\documentclass[preprint,aps,showpacs]{revtex4}

\usepackage{graphicx}
\usepackage{bm}       % bold math
\usepackage{mathptmx} % postscript fonts

\begin{document}
%\hfill BNL-xxx       % for bnl number

\title{Spontaneous violation of CP symmetry in the strong interactions}
\author{Michael Creutz}
\affiliation{
Physics Department, Brookhaven National Laboratory\\
Upton, NY 11973, USA
}
%\email{creutz@bnl.gov}

%\date

\begin{abstract}
{ Some time ago Dashen \cite{dashen} pointed out that spontaneous CP
violation can occur in the strong interactions.  I show how a simple
effective Lagrangian exposes the remarkably large domain of quark mass
parameters for which this occurs.  I close with some warnings for
lattice simulations.}
\end{abstract}

\pacs{
11.30.Er, 12.39.Fe, 11.15.Ha, 11.10.Gh
}% PACS, the Physics and Astronomy
                             % Classification Scheme.
\maketitle

The SU(3) non-Abelian gauge theory of the strong interactions is quite
remarkable in that, once an arbitrary overall scale is fixed, the only
parameters are the quark masses.  Using only a few pseudo-scalar meson
masses to fix these parameters, the non-Abelian gauge theory
describing quark confining dynamics is unique.  It has been known for
some time \cite{dashen} that, as these parameters are varied from
their physical values, exotic phenomena can occur, including
spontaneous breakdown of CP symmetry.  Here I consider the theory with
three flavors of quark and map out in detail the regions of parameter
space where this breaking occurs.

While the spontaneous breaking considered here only occurs in
unphysical regions of parameter space, there are several reasons the
phenomenon may be of wider interest.  Indeed, CP is broken in the real
world, and thus some mechanism along these lines may be useful for
going beyond the strong interactions.  Also, the analysis demonstrates
that when the other quarks are massive, nothing special happens at
vanishing up quark mass.  This raises the question of whether a
non-degenerate massless quark is a physical concept, and is the main
subject of a separate recent paper \cite{Creutz:2003xc}.  These
observations also raise questions for practical lattice calculations
of hadronic physics, where current algorithms ignore any phases in the
fermion determinant and are unable to explore this phenomenon.

The possibility of a spontaneous CP violation is most easily
demonstrated in terms of an effective chiral Lagrangian.  I will begin
with a brief review of this model with three quarks, namely the up,
down, and strange quarks.  This lays the groundwork for discussion of
the CP violating phase.  I will then briefly discuss how heavier
states, most particularly the $\eta^\prime$ meson, enter without
qualitatively changing the picture.  Finally, I make some concluding
remarks on possible impacts of the CP violating structures for lattice
gauge simulations.  An unpublished preliminary version of thise
arguments appears as part of Ref.~\cite{Creutz:2003cj}.  The occurance
of this phenomenon with three degenerate quarks is presented in
Ref.~\cite{Montvay:1999gn}.  A discussion of the CP violating
phenomenon in terms of the analytic structure of the partition
function is in Ref.~\cite{Akemann:2001ir}.

I consider the three flavor theory with its approximate SU(3)
symmetry.  Using three flavors simplifies the discussion, although the
CP violating phase can also be demonstrated for the two flavor theory
following the discussion in Ref.~\cite{mymasspaper}.  I work with the
familiar octet of light pseudo-scalar meson fields $\pi_\alpha$ with
$\alpha=1\ldots8$.  In a standard way (see for example Ref.~\cite{km})
I consider an effective field theory defined in terms of the SU(3)
valued group element
\begin{equation}
\label{sigma}
\Sigma=\exp(i\pi_\alpha \lambda_\alpha/f_\pi)\in SU(3).
\end{equation}
Here the $\lambda_\alpha$ are the usual Gell-Mann matrices, and
$f_\pi$ has a phenomenological value of about 93 MeV.  I follow the
normalization convention that ${\rm Tr} \lambda_\alpha \lambda_\beta =
2\delta_{\alpha\beta}$.  In the chiral limit of vanishing quark
masses, the interactions of the eight massless Goldstone bosons are
modeled with the effective Lagrangian density
\begin{equation}
\label{kinetic}
L_0={f_\pi^2\over 4}{\rm Tr}(\partial_\mu \Sigma^\dagger \partial_\mu \Sigma).
\end{equation}
The non-linear constraint of $\Sigma$ onto the group SU(3) makes this
theory non-renormalizable.  It is to be understood only as the
starting point for an expansion of particle interactions in powers of
their masses and momenta.  Expanding Eq.~(\ref{kinetic}) to second
order in the meson fields gives the conventional kinetic terms for our
eight mesons.

This theory is invariant under parity and charge conjugation,
manifested by
\begin{equation}
P:\ \Sigma\ \rightarrow\ \Sigma^{-1}
\qquad\qquad 
CP:\ \Sigma\ \rightarrow\ \Sigma^{*}
\end{equation}
where the operation $*$ refers to complex conjugation.  The eight
meson fields are pseudo-scalars.  The neutral pion and the eta meson
are both even under charge conjugation.

With massless quarks, the underlying quark-gluon theory has a chiral
symmetry under
\begin{equation}
\psi_L\rightarrow \psi_L g_L
\qquad\qquad
\psi_R\rightarrow \psi_R g_R.
\end{equation}   
Here $(g_L,g_R)$ is in $(SU(3)\times SU(3))$ and $\psi_{L,R}$
represent the chiral components of the quark fields, with flavor
indices understood.  This symmetry is expected to break spontaneously
to a vector SU(3) via a vacuum expectation value for $\overline \psi_L
\psi_R$.  This motivates the sigma model through the identification
\begin{equation}
\langle 0 | \overline \psi_L \psi_R | 0 \rangle \leftrightarrow v \Sigma.
\label{vec}
\end{equation}
The quantity $v$ characterizes the strength of the spontaneous
breaking.  The effective field transforms under the chiral symmetry as
\begin{equation}
\Sigma\rightarrow g_L^\dagger \Sigma g_R.
\end{equation} 
Eq.~(\ref{kinetic}) represents the simplest non-trivial expression
invariant under this symmetry.

Quark masses break the chiral symmetry explicitly.  These are
introduced through a 3 by 3 mass matrix $M$ appearing in an added
potential term
\begin{equation}
L= L_0 - v {\rm Re\ Tr}(\Sigma M).
\end{equation} 
Here $v$ is the same dimensionful factor appearing in Eq.~(\ref{vec}).
The chiral symmetry of our starting theory shows the physical
equivalence of a given mass matrix $M$ with a rotated matrix
$g_R^\dagger M g_L$.  Using this freedom to put the mass matrix into a
standard form, I take it as diagonal with increasing eigenvalues
\begin{equation}
M=\pmatrix{ 
m_u & 0 & 0 \cr
0 & m_d & 0 \cr
0 & 0 & m_s \cr
}
\end{equation} 
representing the up, down, and strange quark masses.

In general the mass matrix can still be complex.  The chiral symmetry
allows one to move phases between the masses, but the determinant of
$M$ is invariant.  Under charge conjugation the mass term would only
be invariant if $M=M^*$.  If $|M|$ is not real, then its phase is the
famous CP violating parameter usually associated with topological
structure in the gauge fields.  Here I take all quark masses as real.
Since I am looking for spontaneous symmetry breaking, I consider the
case where there is no explicit CP violation.

Expanding the mass term quadratically in the meson fields generates
the effective mass matrix for the eight mesons
\begin{equation}
{\cal M}_{\alpha\beta}\ \propto\ {\rm Re\ Tr}\ \lambda_\alpha\lambda_\beta M.
\end{equation}
The isospin-breaking up-down mass-difference plays a crucial role in
the later discussion.  This gives this matrix an off diagonal piece
mixing the $\pi_0$ and the $\eta$
\begin{equation}
{\cal M}_{3,8}\ \propto\ m_u-m_d.
\end{equation}
The eigenvalues of ${\cal M}$ give the standard mass relations
\begin{equation}
\label{mesons}
\matrix{
m_{\pi_+}^2= \  m_{\pi_-}^2\propto m_u+m_d \hfill\cr
m_{K_+}^2= \ m_{K_-}^2\propto m_u+m_s \hfill\cr
m_{K_0}^2= \ m_{\overline K_0}^2\propto m_d+m_s\hfill \cr
m_{\pi_0}^2 \propto\  {2\over 3} \bigg(m_u+m_d+m_s
-\sqrt{m_u^2+m_d^2+m_s^2-m_um_d-m_um_s-m_dm_s}\bigg)\cr
m_{\eta_{\phantom{0}}}^2 \propto \ {2\over 3} \bigg(m_u+m_d+m_s
+\sqrt{m_u^2+m_d^2+m_s^2-m_um_d-m_um_s-m_dm_s}\bigg).\cr
}
\end{equation}  
Here I label the mesons with their conventional names.  From these
relations, ratios of meson masses give estimates for the ratios of the
quark masses \cite{km,weinberg,leutwyler}.

So far all this is standard.  Now I vary the quark masses and look for
interesting phenomena.  In particular, I want to find spontaneous
breaking of the CP symmetry.  Normally the $\Sigma$ field fluctuates
around the identity in SU(3).  However, for some values of the quark
masses this ceases to be true.  When the vacuum expectation of
$\Sigma$ deviates from the identity, some of the meson fields acquire
expectation values.  As they are pseudo-scalars, this necessarily
involves a breakdown of parity, as noted by Dashen \cite{dashen}.

To explore this possibility, concentrate on the lightest meson from
Eq.~(\ref{mesons}), the $\pi_0$.  From Eq.~(\ref{mesons}) one can
calculate the product of the $\pi_0$ and $\eta$ masses
\begin{equation}
\label{prod}
m_{\pi_0}^2m_\eta^2
\ \propto\ m_um_d+m_um_s+m_dm_s.
\end{equation}
The $\pi_0$ mass vanishes whenever
\begin{equation}
\label{boundary}
m_u={-m_sm_d\over m_s+m_d}.
\end{equation}
For increasingly negative up-quark masses, the expansion around
vanishing pseudo-scalar meson fields fails.  The vacuum is no longer
approximated by fluctuations of $\Sigma$ around the unit matrix;
instead it fluctuates about an SU(3) matrix of form
\begin{equation}
\Sigma=\pmatrix{
e^{i\phi_1}&0&0\cr
0&e^{i\phi_2}&0\cr
0&0&e^{-i\phi_1-i\phi_2}\cr
}
\end{equation}
where the phases satisfy
\begin{equation}
m_u \sin(\phi_1)=m_d\sin(\phi_2)=-m_s\sin(\phi_1+\phi_2).
\end{equation}
There are two minimum action solutions, differing by flipping the
signs of these angles.  The transition is a continuous one, with
$\Sigma$ going smoothly to the identity on approaching the boundary in
Eq.~(\ref{boundary}).  The magnitude of these angles controls the
magnitude of the resulting CP violation.

\begin{figure*}
\centering
\includegraphics[width=3in]{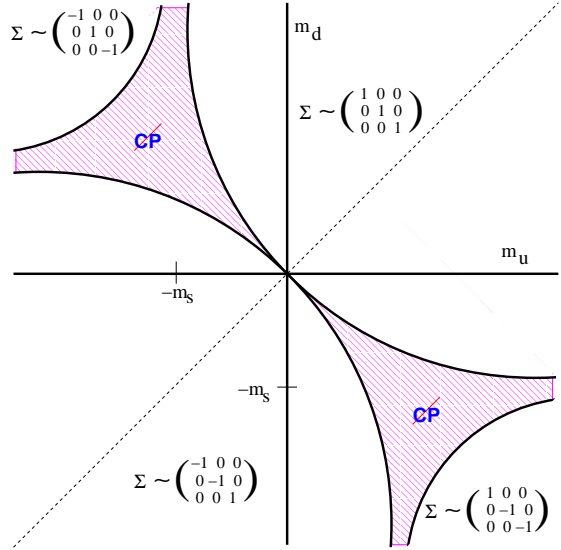}
\caption{\label{fig:phasediagram} The schematic phase diagram of
quark-gluon dynamics as a function of the two lightest quark masses.
The shaded regions exhibit spontaneous CP breaking.  On the diagonal
line with $m_u= m_d$ there are three degenerate pions due to
isospin symmetry.  The neutral pion mass vanishes on the boundary of
the CP violating phase.  The asymptotes of the boundaries are given by
the strange quark mass.  In the CP conserving phases, the vacuum
fluctuates about the indicated values for $\Sigma$.}
\end{figure*}

In the new vacuum the neutral pseudo-scalar meson fields acquire
expectation values.  As they are CP odd, this symmetry is
spontaneously broken.  This will have various experimental
consequences, for example eta decay into two pions becomes allowed
since a virtual third pion can be absorbed by the vacuum.
Fig.~(\ref{fig:phasediagram}) sketches the inferred phase diagram as a
function of the up and down quark masses.  

Chiral rotations insure a symmetry under the flipping of the signs of
both quark masses.  This produces a distinct CP conserving phase.
When the magnitudes of both the up and down quark masses exceed the
strange quark mass, two additional CP conserving phases are found.
The figure indicates the values of $\Sigma$ around which the vacua
fluctuate for the four respective CP conserving phases.

The asymptotes of the boundaries of the CP violating region are
determined by the strange quark mass.  If the strange quark mass is
taken to a large value, then this scale will instead be controlled by
the strong interaction scale.

At first sight the appearance of the CP violating phase at negative up
quark mass may seem surprising.  Naively in perturbation theory the
sign of a fermion mass can be rotated away by a redefinition
$\psi\rightarrow \gamma_5 \psi$.  However this rotation is anomalous,
making the sign of the quark mass observable.  A more general complex
phase in the mass would also have physical consequences, i.e.
explicit CP violation.  With real quark masses the underlying
Lagrangian is CP invariant, but the above discussion shows that there
exists a large region where the ground state spontaneously breaks this
symmetry.

Vafa and Witten \cite{vafawitten} argued on rather general conditions
that CP could not be spontaneously broken in the strong interactions.
However their argument makes positivity assumptions on the path
integral measure.  When a quark mass is negative, the fermion
determinant need not be positive for all gauge configurations; in this
case their assumptions fail.  Azcoiti and Galante \cite{Azcoiti:1999rq}
have also criticized the generality of the Vafa and Witten result.

The possible existence of this phase was anticipated some time ago on
the lattice by Aoki \cite{aoki}.  For the one flavor case he found
this parity breaking phase with Wilson lattice gauge fermions.  He
went on to discuss also two flavors, finding both flavor and parity
symmetry breaking.  The latter case is now regarded as a lattice
artifact.  The chiral breaking terms in the Wilson action open up the
CP violating phase for a finite region along $m_u=m_d$ line. For a
review of these issues see Ref.~\cite{myreview}.

In conventional discussions of CP non-invariance in the strong
interactions \cite{witten} appears a complex phase $e^{i\theta}$
appearing on tunneling between topologically distinct gauge field
configurations.  The famous U(1) anomaly formally allows moving this
phase into the determinant of the quark mass matrix.  Rotating all
phases into the up-quark mass shows that the spontaneous breaking of
CP is occurring at an angle $\theta=\pi$.  Note that when the down
quark mass is positive, the CP violating phase does not appear for
up-quark masses greater than a negative minimum value.  There exists a
finite gap with $\theta=\pi$ without this symmetry breaking.  The
chiral model predicts a smooth behavior in all physical processes as
the up-quark mass passes through zero.

An interesting special case occurs when the up and down quarks have
the same magnitude but opposite sign for their masses, i.e.
$m_u=-m_d$.  In this situation it is illuminating to rotate the minus
sign into the phase of the strange quark.  Then the up and down quark
are degenerate, and an exact vector $SU(2)$ flavor symmetry is
restored.  The spectrum will show three degenerate pions.

The above discussion was entirely in terms of the pseudo-scalar mesons
that become Goldstone bosons in the chiral limit.  One might wonder
how higher states can influence this phase structure.  Of particular
concern is the $\eta^\prime$ meson associated with the anomalous
$U(1)$ symmetry present in the classical quark-gluon Lagrangian.
Non-perturbative processes, including topologically non-trivial gauge
field configurations, are well known to generate a mass for this
particle.  I will now argue that, while this state can shift masses
due to mixing with the lighter mesons, it does not make a qualitative
difference in the existence of a phase with spontaneous CP violation.

The easiest way to introduce the $\eta^\prime$ into the effective
theory is to promote the group element $\Sigma$ to an element of
$U(3)$ via an overall phase factor.  Thus I generalize
Eq.~(\ref{sigma}) to
\begin{equation}
\Sigma=\exp\left(i\pi_\alpha \lambda_\alpha/f_\pi
+i\sqrt{ 2\over 3}\eta^\prime/f_\pi\right)
\in U(3).
\end{equation}
The factor $\sqrt{2/3}$ gives the $\eta^\prime$ field the same
normalization as the $\pi$ fields.  Our starting kinetic Lagrangian in
Eq.~(\ref{kinetic}) would have this particle also be massless.  One
way to fix this deficiency is to mimic the anomaly with a term
proportional to the determinant of $\Sigma$
\begin{equation}
L_0={f_\pi^2\over 4}{\rm Tr}(\partial_\mu \Sigma^\dagger \partial_\mu \Sigma)
-C|\Sigma|.
\end{equation}
The parameter $C$ parameterizes the strength of the anomaly in the
$U(1)$ factor.

On including the mass term exactly as before, additional mixing occurs
between the $\eta^\prime$, the $\pi_0$, and the $\eta$.  The
corresponding mixing matrix takes the form
\begin{equation}
\pmatrix{
m_u+m_d & {m_u-m_d \over \sqrt 3} & \sqrt{2\over 3}(m_u-m_d) \cr
{m_u-m_d \over \sqrt 3} & {m_u+m_d+4m_s \over 3} 
          &  {\sqrt 2(m_u+m_d-2m_s)\over 3}\cr
 \sqrt{2\over 3}( m_u-m_d)&  {\sqrt2 (m_u+m_d-2m_s)\over 3} & {2\over 3}m_a \cr
}
\end{equation}
where $m_a$ characterizes the contribution of the non-perturbative
physics to the $\eta^\prime$ mass.  This should have a value of order
the strong interaction scale; in particular, it should be large
compared to at least the up and down quark masses.  The two by two
matrix in the upper left of this expression is exactly what is
diagonalized to find the neutral pion and eta masses in
Eq.~(\ref{mesons}).

The boundary of the CP violating phase occurs where the determinant of
this matrix vanishes.  This modifies Eq.~(\ref{prod}) to
\begin{equation}
m_{\pi_0}^2m_\eta^2 m_{\eta^\prime}^2
\ \propto\ 
m_a (m_um_d+m_um_s+m_dm_s)-
m_u(m_d-m_s)^2-m_d(m_u-m_s)^2-m_s(m_u-m_d)^2.
\end{equation}
The boundary shifts slightly from the earlier result, but still passes
through the origin, leaving Fig.~(\ref{fig:phasediagram})
qualitatively unchanged.

While I have been exploring rather unphysical regions in parameter
space, these observations do raise some wider issues.  For practical
lattice calculations of hadronic physics, current simulations are done
at relatively heavy values for the quark masses.  This is because the
known fermion algorithms tend to converge rather slowly at light quark
masses.  Extrapolations by several tens of MeV are needed to reach
physical quark masses, and these extrapolations tend to be made in the
context of chiral perturbation theory.  While certainly not a proof of
a problem, the presence of a CP violating phase quite near the
physical values for the quark masses suggests strong variations in the
vacuum state with rather small changes in the up-quark mass; indeed,
less than a 10 MeV change in the traditionally determined up-quark
mass can drastically change the low energy spectrum.  Most simulations
consider degenerate quarks, and chiral extrapolations so far have been
quite successful.  But some quantities, namely certain baryonic
properties \cite{thomas}, do seem to require rather strong variations
as the chiral limit is approached.  These effects and the strong
dependence on the up-quark mass may be related.

Another worrying issue is the validity of current simulation
algorithms with non-degenerate quarks.  With an even number of
degenerate flavors the fermion determinant is positive and can
contribute to a measure for Monte Carlo simulations.  With light
non-degenerate quarks the positivity of this determinant is not
guaranteed.  Indeed, the CP violation can occur only when the fermions
contribute large phases to the path integral.  Current algorithms for
dealing with non-degenerate quarks \cite{milc} take a root of the
determinant with multiple flavors.  In this process any possible
phases are ignored.  Such an algorithm is incapable of seeing the CP
violating phenomena discussed here.  This point may not be too serious
in practice since the up and down quarks are nearly degenerate and the
strange quark is fairly heavy.  Again this is not a proof, but these
issues should serve as a warning that things might not work as well as
desired.

\section*{Acknowledgements}
This manuscript has been authored under contract number
DE-AC02-98CH10886 with the U.S.~Department of Energy.  Accordingly,
the U.S. Government retains a non-exclusive, royalty-free license to
publish or reproduce the published form of this contribution, or allow
others to do so, for U.S.~Government purposes.

\end{document}